\documentclass[AMA,STIX1COL]{WileyNJD-v2}
\usepackage{dsfont}
\usepackage[T1]{fontenc}
\usepackage[utf8]{inputenc}
\usepackage{bm}

\PassOptionsToPackage{unicode}{hyperref}
\PassOptionsToPackage{hyphens}{url}

\usepackage{ifxetex,ifluatex}
\ifnum 0\ifxetex 1\fi\ifluatex 1\fi=0 
  \usepackage[T1]{fontenc}
  \usepackage[utf8]{inputenc}
\else 
  \usepackage{unicode-math}
  \defaultfontfeatures{Scale=MatchLowercase}
  \defaultfontfeatures[\rmfamily]{Ligatures=TeX,Scale=1}
\fi
\IfFileExists{upquote.sty}{\usepackage{upquote}}{}
\IfFileExists{microtype.sty}{
  \usepackage[]{microtype}
  \UseMicrotypeSet[protrusion]{basicmath} 
}{}
\usepackage{xcolor}
\IfFileExists{xurl.sty}{\usepackage{xurl}}{} 
\IfFileExists{bookmark.sty}{\usepackage{bookmark}}{\usepackage{hyperref}}
\hypersetup{
  pdfauthor={Antonio Remiro-Azócar},
  hidelinks,
  pdfcreator={LaTeX via pandoc}}
\usepackage{fancyvrb}

\DefineVerbatimEnvironment{Highlighting}{Verbatim}{commandchars=\\\{\}}
\usepackage{framed}
\definecolor{shadecolor}{RGB}{248,248,248}
\newenvironment{Shaded}{\begin{snugshade}}{\end{snugshade}}

\newcommand{\CommentTok}[1]{\textcolor[rgb]{0.56,0.35,0.01}{\textit{#1}}}

\newcommand{\ControlFlowTok}[1]{\textcolor[rgb]{0.13,0.29,0.53}{\textbf{#1}}}
\newcommand{\DataTypeTok}[1]{\textcolor[rgb]{0.13,0.29,0.53}{#1}}
\newcommand{\DecValTok}[1]{\textcolor[rgb]{0.00,0.00,0.81}{#1}}

\newcommand{\FloatTok}[1]{\textcolor[rgb]{0.00,0.00,0.81}{#1}}

\newcommand{\KeywordTok}[1]{\textcolor[rgb]{0.13,0.29,0.53}{\textbf{#1}}}
\newcommand{\NormalTok}[1]{#1}
\newcommand{\OperatorTok}[1]{\textcolor[rgb]{0.81,0.36,0.00}{\textbf{#1}}}
\newcommand{\OtherTok}[1]{\textcolor[rgb]{0.56,0.35,0.01}{#1}}

\newcommand{\StringTok}[1]{\textcolor[rgb]{0.31,0.60,0.02}{#1}}

\usepackage{graphicx}
\makeatletter
\def\maxwidth{\ifdim\Gin@nat@width>\linewidth\linewidth\else\Gin@nat@width\fi}
\def\maxheight{\ifdim\Gin@nat@height>\textheight\textheight\else\Gin@nat@height\fi}
\makeatother
\setkeys{Gin}{width=\maxwidth,height=\maxheight,keepaspectratio}
\makeatletter
\def\fps@figure{htbp}
\makeatother
\articletype{Commentary}%

\received{26 April 2016}
\revised{6 June 2016}
\accepted{6 June 2016}

\usepackage{booktabs}
\usepackage{pifont}

\usepackage{textcomp}

\makeatletter

\newcommand\midtilde@raisedtilde[1][.5]{\raisebox{#1ex}{\texttildelow}}
\def\midtilde@normaltilde{\texttildelow}

\newcommand\midtilde%
{%
  \expandafter\in@\expandafter{\f@family}%
    {cmr,cmss,cmtt,cmm,cmsy,cmx,%
    lmr,lmss,lmtt,lmm,lmsy,lmx,%
    pxr,pxss,pxm,pxsy,pxx,%
    txr,txss,txm,txsy,txx}%
  \ifin@%
    \midtilde@raisedtilde%
  \else%
    \expandafter\in@\expandafter{\f@family}%
    {pxtt,txtt}%
    \ifin@%
      \midtilde@raisedtilde[.35]%
    \else%
      \midtilde@normaltilde%
    \fi%
  \fi%
}

\begin{document}

\title{Effect modification in anchored indirect treatment comparisons: Comments on “Matching-adjusted indirect comparisons: Application to time-to-event data”}

\author[1,2]{Antonio Remiro-Az\'ocar}

\author[1,3,4]{Anna Heath}

\author[1]{Gianluca Baio}

\authormark{REMIRO-AZ\'OCAR \textsc{et al}}

\address[1]{\orgdiv{Department of Statistical Science}, \orgname{University College London}, \orgaddress{\state{London}, \country{United Kingdom}}}

\address[2]{\orgdiv{Medical Affairs Statistics}, \orgname{Bayer plc}, \orgaddress{\state{Reading}, \country{United Kingdom}}}

\address[3]{\orgdiv{Child Health Evaluative Sciences}, \orgname{The Hospital for Sick Children}, \orgaddress{\state{Toronto}, \country{Canada}}}

\address[4]{\orgdiv{Dalla Lana School of Public Health}, \orgname{University of Toronto}, \orgaddress{\state{Toronto}, \country{Canada}}}

\corres{*Antonio Remiro Az\'ocar, Department of Statistical Science, University College London, London, United Kingdom. \email{antonio.remiro.16@ucl.ac.uk}. Tel: (+44 20) 7679 1872. Fax: (+44 20) 3108 3105}

\presentaddress{Antonio Remiro Az\'ocar, Department of Statistical Science, University College London, Gower Street, London, WC1E 6BT, United Kingdom}

\abstract{This commentary regards a recent simulation study conducted by Aouni, Gaudel-Dedieu and Sebastien, evaluating the performance of different versions of matching-adjusted indirect comparison (MAIC) in an anchored scenario with a common comparator. The simulation study uses survival outcomes and the Cox proportional hazards regression as the outcome model. It concludes that using the LASSO for variable selection is preferable to balancing a maximal set of covariates.  However, there are no treatment effect modifiers in imbalance in the study. The LASSO is more efficient because it selects a subset of the maximal set of covariates but there are no cross-study imbalances in effect modifiers inducing bias. We highlight the following points: (1) in the anchored setting, MAIC is necessary where there are cross-trial imbalances in effect modifiers; (2) the standard indirect comparison provides greater precision and accuracy than MAIC if there are no effect modifiers in imbalance; (3) while the target estimand of the simulation study is a conditional treatment effect, MAIC targets a marginal or population-average treatment effect; (4) in MAIC, variable selection is a problem of low dimensionality and sparsity-inducing methods like the LASSO may be problematic. Finally, data-driven approaches do not obviate the necessity for subject matter knowledge when selecting effect modifiers. R code is provided in the Appendix to replicate the analyses and illustrate our points.}


\maketitle

\renewcommand{\thefootnote}{\alph{footnote}}


We read with interest the recent paper by Aouni, Gaudel-Dedieu and Sebastien\cite{aouni2021matching} (AGS), in which they examine the statistical performance of matching-adjusted indirect comparison (MAIC) in an ``anchored'' scenario with a common comparator treatment across studies. MAIC is a relatively novel methodology used to compare the efficacy or effectiveness of treatments in a two-study scenario, often seen in health technology assessment, where there is access to individual patient data (IPD) on covariates and outcomes from one study, and only access to published aggregate-level data (ALD) --- typically a limited set of covariate moments (``Table 1'') and summary outcomes --- from the other study. AGS denote the IPD study, comparing intervention $A$ to common comparator $C$, as ``study A'' (population $S=1$), and the ALD study, comparing intervention $B$ to common comparator $C$ as ``study B'' (population $S=2$). In this sparse network, with only one study per active treatment, indirect comparisons are vulnerable to bias induced by cross-trial differences in baseline covariates. 


AGS compare several implementations of MAIC in a simulation study, using time-to-event outcomes and the Cox proportional hazards regression as the outcome model. AGS conclude that using the LASSO technique ``is better'' than balancing a maximal set of covariates. We raise the following points about AGS' conclusions: 

\begin{enumerate}
    \item In the anchored setting, MAIC is necessary where there are cross-study imbalances in effect modifiers that induce bias. If this is the case, MAIC requires accounting for \textit{all} effect modifiers in order to satisfy the \textit{conditional constancy of relative effects} assumption and be unbiased.\cite{phillippo2018methods, phillippo2016nice}
    \item In the data-generating mechanism of the simulation study, there is no treatment effect modification. Therefore, there is no bias to remove and accuracy is driven entirely by precision. LASSO has lower variance because it selects a subset of the maximal set of covariates. In the simulation study setting, MAIC is not necessary --- the standard indirect comparison would provide the greatest precision and accuracy. 
    \item The target estimands of the simulation study are conditional treatment effects (a ratio of conditional hazard ratios) but MAIC targets marginal or population-average treatment effects (a ratio of marginal hazard ratios). These measures of effect may not coincide for the hazard ratio, even under covariate balance and no confounding, as the measure of effect is non-collapsible. The bias observed is potentially due to this conflation of effects.
    \item In the context of MAIC, variable selection is typically a problem of low dimensionality. A sparsity-inducing method like LASSO that shrinks the regression coefficients may be problematic and induce bias, due to breaking the conditional constancy of relative effects. Data-driven approaches do not obviate the necessity for substantive subject matter knowledge when selecting effect modifiers.   
\end{enumerate}

\noindent These points are discussed in more detail in the below paragraphs. In the Appendix, we attempt to replicate the analyses of AGS using the information provided in the original article and illustrate some of the aforementioned points. 

\section*{Effect modification and method assumptions}

\subsection*{Effect modification}

Using the notation of AGS, covariate $X$ is an \textit{effect measure modifier} (\textit{effect modifier} for short) if the effect of treatment $T$ on outcome $\mathcal{O}$, on a specific scale, varies by level or strata of $X$.\cite{vanderweele2007four}

Studies A and B are both randomized controlled trials (RCTs). In an ideal or perfectly-executed scenario with perfect measurement and no missing data, the nature of randomization implies that, in expectation, there is covariate balance (i.e., the covariates are similarly distributed) and the active treatment and control groups are exchangeable. The random assignment of units to treatment eliminates confounding by design,\cite{pocock2013clinical,friedman2015fundamentals} providing protection from the confounding of the treatment effect by baseline covariates. Hence, ``within-study'' inferences are internally valid — one can produce unbiased estimates for the marginal $A$ vs. $C$ treatment effect in the $S=1$ population (we shall denote this estimand $\Delta_{AC}^{(1)}$), and for the marginal $B$ vs. $C$ treatment effect in the $S=2$ population (we denote this estimand $\Delta_{BC}^{(2)}$).

In the context of study A, different levels of the effect modifiers of treatment $A$ are associated with differential marginal or population-average treatment effects for $A$ vs. $C$. Hence, the set of effect modifiers is the source of treatment effect variation, fully explaining the heterogeneity of the effect of treatment $A$ vs. $C$. If there are differences in effect modifiers between the $S=1$ population and the $S=2$ population, implicitly assumed to be the target population by MAIC, the marginal treatment effect for $A$ vs. $C$ in $S=1$ is expected to differ from the marginal treatment effect for $A$ vs. $C$ in $S=2$.\cite{weisberg2009selection, cole2010generalizing, olsen2013external, dahabreh2016using, stuart2015assessing, lesko2017generalizing} Namely, if covariate $X$ modifies the effect of treatment $A$, i.e., such treatment effect is heterogeneous, and the distribution of the factor $X$ in $S=1$ differs from the distribution of that factor in $S=2$, we typically have $\Delta_{AC}^{(1)} \neq \Delta_{AC}^{(2)}$.\footnote{Except in the pathological case where the bias induced by different effect modifiers is in opposite directions and cancels out.} The difference between $\Delta_{AC}^{(1)}$ and $\Delta_{AC}^{(2)}$ is a function of the treatment effect heterogeneity induced by the effect modifiers and the imbalance (difference in means) in effect modifiers across populations. This heterogeneity is the primary motivation for the use of MAIC in the anchored setting. 

\subsection*{Method assumptions}

Standard indirect comparison methods such as the Bucher method\cite{bucher1997results} do not explicitly produce an estimate for any target population in particular --- it is not typically stated whether the target population is $S=1$, $S=2$ or otherwise, regardless of whether the analysis is based on ALD or on IPD from each study.\cite{manski2019meta} The Bucher method estimates the marginal treatment effect for $A$ vs. $B$ as:
\begin{equation*}
\hat{\Delta}_{AB} = \hat{\Delta}_{AC} - \hat{\Delta}_{BC},
\end{equation*}
where $\hat{\Delta}_{AC}$ is the estimated marginal treatment effect of $A$ vs. $C$, and $\hat{\Delta}_{BC}$ is the estimated marginal treatment effect of $B$ vs. $C$. This comparison is only valid for any target population where treatment effects are not heterogeneous or all potential effect modifiers (for $A$ vs. $C$ and $B$ vs. $C$) are equidistributed across studies. 

In the context of the article by AGS, one only has access to published ALD for study B. Therefore, the indirect comparison is conducted by necessity in population $S=2$. The Bucher method would estimate the marginal treatment effect for $A$ vs. $B$ in the $S=2$ population as:
\begin{equation*}
\hat{\Delta}_{AB}^{(2)} = \hat{\Delta}_{AC}^{(1)} - \hat{\Delta}_{BC}^{(2)}.
\end{equation*}
Here, $\hat{\Delta}_{AC}^{(1)}$ is the estimated marginal treatment effect of $A$ vs. $C$ in the $S=1$ population, using the IPD for study A, and $\hat{\Delta}_{BC}^{(2)}$ is the estimated marginal treatment effect of $B$ vs. $C$ in $S=2$, available in the published ALD or calculated from published aggregate outcomes. In order for this comparison to be unbiased, a requirement is the \textit{constancy of relative effects}:\cite{phillippo2018methods, phillippo2016nice} $\Delta_{AC}^{(1)} = \Delta_{AC}^{(2)}$. This means that all covariates, measured or unmeasured, modifying the effect of treatment $A$ versus $C$ must be balanced across the populations.\footnote{If ALD were available for study A and IPD for study B, the comparison would have to be conducted in $S=1$, and would require the cross-study balance of all covariates modifying the effect of treatment $B$.}  

The objective of MAIC is to weight the IPD for study A so that it resembles the population of study B ($S=2$), with respect to the distribution of selected baseline covariates. The weighted IPD is then used to estimate the direct target of MAIC:  $\Delta_{AC}^{(2)}$, the marginal $A$ vs. $C$ treatment effect in the $S=2$ population. Then, an adjusted indirect comparison is performed in the $S=2$ population, where the marginal $A$ vs. $B$ treatment effect is estimated as:
\begin{equation*}
\hat{\Delta}_{AB}^{(2)} = \hat{\Delta}_{AC}^{(2)} - \hat{\Delta}_{BC}^{(2)},
\end{equation*}
where $\hat{\Delta}_{AC}^{(2)}$ is the estimated marginal treatment effect of $A$ vs $C$ in $S=2$.

AGS imply that the weighting procedure should include all the variables that explain the \textit{absolute} outcome $\mathcal{O}$ under active treatment $A$. Conditional on these prognostic covariates, the distribution of the absolute outcomes is independent of the population. An important point is that this assumption can be relaxed in the anchored scenario. In an anchored MAIC, a comparison of \textit{relative} outcomes or effects, not absolute outcomes under each treatment, is of interest. An anchored comparison only requires identifying and balancing the effect modifiers, which are the covariates that explain the heterogeneity of the $A$ vs. $C$ treatment effect. This is likely a smaller set of variables than that of variables explaining the absolute outcomes.  

An anchored MAIC assumes the \textit{conditional constancy of relative effects}\cite{phillippo2018methods, phillippo2016nice} across populations. Namely, given the selected effect-modifying covariates, the (weighted) covariate-adjusted marginal treatment effect for $A$ vs. $C$ in population $S=1$ is equal to the unadjusted marginal treatment effect for $A$ vs. $C$ in $S=2$.\footnote{There are many ways to articulate the assumption of conditional constancy of relative effects. Other formulations appear in the ``generalizability'' or ``transportability'' literature.\cite{cole2010generalizing, kern2016assessing,stuart2011use, hartman2015sample} One can consider that being in study A ($S=1$) or study B ($S=2$) does not carry over any information about the marginal $A$ vs $C$ treatment effect, once we condition on the treatment effect modifiers. Namely, we can assume that trial assignment/selection is conditionally ``ignorable'', unconfounded or exchangeable for such marginal treatment effect, i.e., conditionally independent of the treatment effect, given the selected effect modifiers. This means that after adjusting for these effect modifiers, the treatment effect and trial assignment are conditionally independent.} That is, the weighting model needs to include \textit{all} the variables that modify the effect of treatment $A$, whether measured or unmeasured. AGS posit that an ``influential covariate'' can be ignored if it is balanced between the populations. However, one must account for \textit{all} effect modifiers, regardless of balance between studies, because excluding balanced covariates from the weighting procedure does not ensure balance after the weighting.\cite{phillippo2018methods, phillippo2016nice} The MAIC estimate $\hat{\Delta}_{AC}^{(2)}$ for the marginal $A$ vs. $C$ treatment effect in population $S=2$ will be biased if any pre-treatment covariates that modify the treatment effect are unobserved (either in study A or study B) or unaccounted for in the weighting model.

The conditional constancy of relative effects is an untestable assumption because the $A$ vs. $C$ trial has not been conducted in the $S=2$ population, where outcomes under treatment $A$ are unobserved. It is also a very demanding assumption in practice due to a variety of reasons. Firstly, it requires that all effect modifiers of treatment $A$ are measured in the IPD for study A. In addition, the set of published baseline characteristics for study B must be sufficiently rich, such that all effect modifiers of treatment $A$ are available. Unfortunately, this is a key challenge as data on the $S=2$ population are often limited, and it can sometimes be difficult to find common measures between different studies.\cite{stuart2015assessing, stuart2017generalizing} Moreover, only marginal moments of the covariates are published for the ALD study, with data on joint covariate distributions typically unavailable. Marginal balance across study populations does not guarantee multidimensional balance across the full joint distributions without further distributional assumptions.\cite{hong2019comparison} Conditional constancy could be broken if there are higher-order treatment-by-covariate interactions, involving two or more covariates. Importantly, the conditional constancy assumption is tied to the scale used to define the treatment effects and effect modifiers. For a given covariate, treatment effect modification may exist on one scale but not on another.\cite{brumback2008effect} Furthermore, the effect modifier status of a variable can itself be ``modified'' by simultaneously accounting for other variables.\cite{vanderweele2007four} 

\section*{Simulation study protocol}

As explained earlier, the divergence between $\Delta_{AC}^{(1)}$ and $\Delta_{AC}^{(2)}$ is a function of: (1) the extent to which each covariate modifies the treatment effect; and (2) the imbalance in effect modifiers across studies, which is the extent to which the effect modifiers are related to trial assignment. In the simulation study protocol of AGS we face the latter threat but not the former. 

Two covariates, Age and ISS, govern trial assignment through a logistic regression model: $\textnormal{logit}(\mathbb{P}(S=1\vert \textnormal{covariates})) = \theta_1 \times (\textnormal{Age}-65) + \theta_2 \times \textnormal{ISS}$, with the parameter values set to $\theta_1=\theta_2=0.1$. Age and ISS, a binary disease indicator, determine the trial selection mechanism by which subjects come to be assigned to one study or the other. This creates some covariate imbalance between the populations of the two studies. However, while Age and ISS are associated with trial selection, they do not modify the $A$ vs. $C$ marginal treatment effect. Therefore, any covariate imbalances do not induce bias in the standard indirect comparison. 

As there are no covariates modifying the effect of treatment $A$ vs. $C$ in the data-generating mechanism, the marginal treatment effects $\Delta_{AC}^{(1)}$ and $\Delta_{AC}^{(2)}$ are identical. In our Appendix, these are computed as log hazard ratios of $\Delta_{AC}^{(1)}=\log(0.76)$ and $\Delta_{AC}^{(2)}=\log(0.76)$. Because there is no treatment effect variation or heterogeneity, the assumptions of the standard indirect comparison hold. This will be unbiased because the naive IPD estimate $\hat{\Delta}_{AC}^{(1)}$ is an unbiased estimate of $\Delta_{AC}^{(2)}$, and is applicable to the $S=2$ population. MAIC will also be unbiased because there are no effect modifiers to account for. Therefore, no effect modifiers have been unaccounted for, and the MAIC estimate $\hat{\Delta}_{AC}^{(2)}$ is an unbiased estimate of $\Delta_{AC}^{(1)}$ across the simulation scenarios. In this case, the use of a more complex method like MAIC with a larger number of assumptions is not warranted --- weighting induces an increase in variance without the potential for bias reduction. 


These points are illustrated in our Appendix, which can be viewed as a supplement to this section. In the outcome-generating mechanism of AGS, when accounting for different covariate sets in the weighting model, MAIC provides unbiased estimates of $\Delta_{AC}^{(1)}=\log(0.76)$. When we add an imbalanced effect modifier, Age, to the outcome-generating process and account for Age in the weighting model, the MAIC estimate in the $S=2$ population is biased for $\Delta_{AC}^{(1)}$. Similarly, the naive IPD Bucher estimate would be biased for $\Delta_{AC}^{(2)}$. 

\section*{Clarification of estimands}

Further clarification of the treatment effect targeted by the simulation study, and of the estimand targeted by MAIC, is required. Consider the time-to-event setting presented by the original article. Using our own notation, each randomized trial has survival endpoint $Y$, subject to censoring, a dichotomous treatment indicator $T$ (coded 1 for active treatment and 0 for control), and a set of baseline covariates $\bm{X}$ which are prognostic for $Y$. In the simulation study design, the hazard function for each subject at follow-up time $y$ follows the Cox model:
\begin{equation}
h(y \mid \bm{X}, T) = h_0(y) \exp (\boldsymbol{\beta}^\top \bm{X} + \beta_T T),
\label{eqn1}
\end{equation}
where $h_0(\cdot)$ is the baseline hazard function, assumed to be constant, and $\bm{\beta}$ and $\beta_T$ denote coefficients for the prognostic covariates and treatment, respectively. The hazard function above is conditional on $\bm{X}$ and $T$. The \textit{conditional} hazard ratio comparing active to control treatment at $y$ is $h(y \mid \bm{X}, T=1)/h(y \mid \bm{X}, T=0) = \exp{(\beta_T)}$. The \textit{marginal} or \textit{population-average} hazard ratio at $y$ is $h(y \mid T=1)/h(y \mid T=0)$. 

We return to the notation used in the original article (Section 2.3). Because the treatment coefficients in the outcome-generating model, $b_A=\log(0.53)$ and $b_B=\log(0.55)$, AGS state that the ``true hazard ratio of treatment $A$ vs. treatment $B$ is equal to $0.53/0.55=0.964$''. It is worth highlighting that $b_A$ and $b_B$ ($\beta_T$ in Equation \ref{eqn1} above) are conditional treatment effects because they are coefficients of the outcome-generating Cox regression, conditional on the effects of the prognostic variables that have also been included in the model. That is, $b_A$ represents the true average log hazard ratio, at the individual level, of changing a subject's treatment from $C$ to $A$ (the average treatment effect conditioned on the average combination of covariates in $S=1$, or the average effect across sub-populations of subjects who share the same covariates). Similarly, $b_B$ denotes the true average effect in $S=2$, at the unit level, of changing a subject's treatment from $C$ to $B$. 

The target estimands of the simulation study are conditional or unit-level treatment effects. However, MAIC targets an estimand that is calibrated at a different hierarchical level. This estimand is a marginal treatment effect, which has a different interpretation than the conditional treatment effect. The marginal treatment effect is the average effect, at the \textit{population level}, of moving everyone in the population from one treatment to another. MAIC is a method based on propensity score weighting, where the $A$ vs. $C$ treatment effect estimate targets a \textit{marginal} log hazard ratio, rather than a \textit{conditional} log hazard ratio. This is because the outcome model is a univariable weighted regression of outcome on treatment assignment. The estimated treatment effect is the fitted treatment coefficient of this weighted regression. This coefficient estimates a relative effect between subjects that, on expectation, have the same distribution of covariates, corresponding to population $S=2$. 

AGS may have derived the $B$ vs. $C$ treatment effect in $S=2$ by fitting a simple Cox regression of outcome on treatment assignment. This is regularly the case for treatment effects reported in RCT publications, where evidence is often stated at the population level. In that case, the estimated (log) hazard ratio for $B$ vs. $C$ also targets a marginal treatment effect. It is assumed that the objective of methodologies such as MAIC in health technology assessment is to help make inferences and policy decisions at the population level, not the individual level.

The (log) hazard ratio is a non-collapsible measure of effect. Therefore, marginal and conditional (log) hazard ratios may not coincide, even where there is covariate balance and the absence of confounding.\cite{austin2013performance} In this particular simulation study, the observed relative bias (Tables 5 to 9 of the original article by AGS) may be due to a conflation of marginal and conditional measures of effect. While the true target estimand is defined as a ratio of conditional treatment effects (hazard ratios as opposed to log hazard ratios, in this case), MAIC targets a ratio of marginal treatment effects. In our Appendix, the ratio of marginal hazard ratios is computed as 0.984, which does not coincide with the ratio of conditional hazard ratios, $\exp(b_A)/\exp(b_B)=0.53/0.55=0.964$, reported by AGS as the true target estimand. Bias has also been induced in other simulation studies where the measure of effect is non-collapsible, due to targeting the wrong measure of effect.\cite{remiro2020methods, remiro2021conflating} Collapsibility does not hold for most measures of interest in population-adjusted indirect comparisons, such as (log) hazard ratios or (log) odds ratios in oncology applications, when investigating time-to-event or binary outcomes, respectively. 


\section*{Concluding remarks}

Our commentary has some implications for the conclusions of AGS. In the original article, AGS explore use of the LASSO regression for covariate selection. This approach is perceived to be ``better'' and ``more efficient'' than selecting a maximal set of covariates because this ``reduces the bias as well as the SE''. 

In our opinion, the bias for both approaches is about the same in the simulation study. This is because marginal effects are constant across studies --- any implementation of MAIC is likely unbiased, as is the standard indirect treatment comparison. Secondly, LASSO has greater precision because it will select a subset of the maximal set of covariates. As none of the covariates are effect modifiers, including more of these in the weighting model does not remove bias further but reduces the effective sample size and inflates the standard error. This may explain why the LASSO is more accurate or efficient. The standard indirect comparison would be even more accurate in these simulation settings because it does not adjust for any covariates. 

These results confirm that covariates that do not modify the $A$ vs. $C$ treatment effect should be excluded from an anchored MAIC, as recommended by current guidance.\cite{phillippo2018methods, phillippo2016nice} There is a disadvantage to unnecessary or excessive adjustment. Considering bias-variance trade-offs, accounting for a covariate that is not an effect modifier is a sub-optimal use of information because it increases variance and affects the precision of the treatment effect estimate negatively. Similarly, failure to include an effect-modifying covariate is also a sub-optimal use of information that results in increased bias. 

Further research should evaluate the performance of the LASSO in a scenario where there are cross-study imbalances in effect modifiers inducing bias. Nevertheless, we highlight two caveats that are important to bear in mind:

\begin{enumerate}
\item \textbf{Sparsity-inducing methods.} The LASSO imposes a penalty which shrinks all regression coefficients towards zero and directly sets some to zero, thereby providing a variable selection procedure. While AGS perceive this to be an advantage, it may be a disadvantage in the typical application of MAIC, where the covariate selection problem is one of low dimensionality. In practice, there may be little comparability of measures across data sources. We are likely only capable of selecting a subset of the true effect modifiers because we only have access to a few covariates from the published ALD from study B. Anchored MAIC is susceptible to bias where effect modifiers are omitted because the conditional constancy of relative effects is invalid. Therefore, use of a sparsity-promoting method may be problematic; unless there is a large number of potential effect modifiers and poor overlap inducing imprecision (outweighing the potential for bias reduction). 

\item \textbf{Interaction testing.} Within the biostatistics literature, effect modification is usually referred to as interaction, because effect modifiers are considered to alter the effect of treatment by interacting with it on a specific scale,\cite{vanderweele2009concerning} and are often identified by evaluating statistical interaction terms in regression models fitted to the IPD.\cite{phillippo2016nice, rothman1980concepts, simon1982patient, vanderweele2019principles, maldonado1993simulation} While the standard LASSO does not include interactions and is restricted to look for main linear effects only,\cite{tibshirani1996regression} there are extensions specifically designed to consider interaction terms.\cite{lim2015learning, shah2016modelling} Nevertheless, any data-driven approach based on statistical testing will be hindered by the relatively modest sample sizes of individual RCTs. These are, almost invariably, underpowered to assess treatment effect heterogeneity,\cite{peterson1993sample, fisher2017meta} with statistical tests only detecting very large and possibly implausible interactions.\cite{greenland1983tests} Meta-analyses of multiple trials, using IPD or ALD, may provide greater power.\cite{fisher2017meta,dias2018network} 
\end{enumerate}

Ultimately, the selection of effect modifiers to include in the weighting model requires thorough care and investigation. Treatment effect heterogeneity is a complex and largely unknown process, likely to require leveraging both statistical and clinical expertise. It may be reasonable to balance a variable if there is a strong biological reasoning for effect modification, even if the interaction is statistically weak. The identification of effect modifiers will likely depend on a combination of data exploration and analysis, prior subject matter knowledge, firm clinical or biological hypotheses, and literature reviews. Novel causal inference methods, typically relying on directed acyclic graphs,\cite{vanderweele2019principles, ferguson2020evidence} also show promise, but note that these themselves are often based upon background knowledge and expert opinion. Admittedly, strong theory on treatment effect modification is often not available, particularly for novel therapies with unknown mechanisms of action. Finally, the inclusion of all effect modifiers is an unverifiable assumption. Sensitivity analyses are crucial and should be conducted under alternative effect modifier specifications to explore the dependence of inferences on this selection and the robustness of results.\cite{nguyen2017sensitivity, nguyen2018sensitivity}

One of the reviewers raises the problem of valid inference post-model selection with the LASSO\cite{meinshausen2009p} and potential sample-splitting solutions.\cite{meinshausen2009p, wasserman2009high} The LASSO's biased significance tests and the reduced power that comes with sample-splitting are relevant for the statistical detection of interactions. Nevertheless, we highlight that the propensity score model for the weights is a nuisance model fitted before the final step of marginal effect estimation in MAIC. We do not necessarily seek valid inferences for the ``pre-processing'' step used to select the model, focusing instead on valid inference in the final step. One can view estimation of the weights or of alternative nuisance models as a ``best prediction'' problem, for which statistical learning or data-driven methods show some potential.\cite{blakely2020reflection, wager2018estimation, zivich2021machine}

\section*{Appendix}\label{Appendix}

In this appendix, we attempt to replicate the analyses by AGS using the information provided in their article. Some of the points discussed in our commentary are exemplified. We perform the analyses using \texttt{R} software version 3.6.3.\cite{team2013r} 

\subsection*{Data-generating mechanism}

We simulate large populations of 100,000 subjects for both $S=1$ (study A) and $S=2$ (study B), with 50,000 units under each treatment. Tables 1 and 2 of the article by AGS report descriptive statistics of the simulated studies, where covariate means across the simulations are included. Not much information is provided on the underlying joint covariate distributions used to generate the study populations, e.g. on the moments (means and standard deviations) and forms of the marginal distributions, and on the correlation structure of the real database, which AGS resample via the bootstrap to simulate the covariates. 

We use the information on overall covariate means, ignoring sampling variability, and make certain parametric assumptions about the covariate distributions to simulate the populations. It is assumed that the studies are appropriately randomized; hence it should make no difference to simulate the arms of each study jointly or separately, using the arm-specific covariate means. We assume that the covariates are uncorrelated.

\begin{Shaded}
\begin{Highlighting}[]
\KeywordTok{rm}\NormalTok{(}\DataTypeTok{list=}\KeywordTok{ls}\NormalTok{())}
\KeywordTok{set.seed}\NormalTok{(}\DecValTok{555}\NormalTok{) }\CommentTok{\# set random seed for reproducibility}

\NormalTok{N <{-}}\StringTok{ }\DecValTok{100000} \CommentTok{\# number of simulated subjects per population}

\NormalTok{trt <{-}}\StringTok{ }\KeywordTok{c}\NormalTok{(}\KeywordTok{rep}\NormalTok{(}\DecValTok{1}\NormalTok{, N}\OperatorTok{/}\DecValTok{2}\NormalTok{), }\KeywordTok{rep}\NormalTok{(}\DecValTok{0}\NormalTok{, N}\OperatorTok{/}\DecValTok{2}\NormalTok{)) }\CommentTok{\# assume a 1:1 treatment allocation ratio}

\CommentTok{\# population S=1 (study A)}
\NormalTok{age\_S1 <{-}}\StringTok{ }\KeywordTok{rnorm}\NormalTok{(}\DataTypeTok{n=}\NormalTok{N, }\DataTypeTok{mean=}\FloatTok{69.3}\NormalTok{, }\DataTypeTok{sd=}\DecValTok{5}\NormalTok{) }\CommentTok{\# assume Age is normally{-}distributed, SD=5}
\NormalTok{plnen\_S1 <{-}}\StringTok{ }\KeywordTok{rpois}\NormalTok{(}\DataTypeTok{n=}\NormalTok{N, }\DataTypeTok{lambda=}\FloatTok{3.4}\NormalTok{) }\CommentTok{\# assume PLNEN is Poisson{-}distributed}
\NormalTok{iss\_S1 <{-}}\StringTok{ }\KeywordTok{rbinom}\NormalTok{(}\DataTypeTok{n=}\NormalTok{N, }\DataTypeTok{size=}\DecValTok{1}\NormalTok{, }\DataTypeTok{prob=}\FloatTok{0.74}\NormalTok{) }\CommentTok{\# assume ISS is Bernoulli{-}distributed}
\NormalTok{refr\_S1 <{-}}\StringTok{ }\KeywordTok{rbinom}\NormalTok{(}\DataTypeTok{n=}\NormalTok{N, }\DataTypeTok{size=}\DecValTok{1}\NormalTok{, }\DataTypeTok{prob=}\FloatTok{0.92}\NormalTok{) }\CommentTok{\# assume Refr is Bernoulli{-}distributed}

\CommentTok{\# population S=2 (study B)}
\NormalTok{age\_S2 <{-}}\StringTok{ }\KeywordTok{rnorm}\NormalTok{(}\DataTypeTok{n=}\NormalTok{N, }\DataTypeTok{mean=}\FloatTok{62.1}\NormalTok{, }\DataTypeTok{sd=}\DecValTok{5}\NormalTok{)}
\NormalTok{plnen\_S2 <{-}}\StringTok{ }\KeywordTok{rpois}\NormalTok{(}\DataTypeTok{n=}\NormalTok{N, }\DataTypeTok{lambda=}\FloatTok{3.4}\NormalTok{)}
\NormalTok{iss\_S2 <{-}}\StringTok{ }\KeywordTok{rbinom}\NormalTok{(}\DataTypeTok{n=}\NormalTok{N, }\DataTypeTok{size=}\DecValTok{1}\NormalTok{, }\DataTypeTok{prob=}\FloatTok{0.77}\NormalTok{)}
\NormalTok{refr\_S2 <{-}}\StringTok{ }\KeywordTok{rbinom}\NormalTok{(}\DataTypeTok{n=}\NormalTok{N, }\DataTypeTok{size=}\DecValTok{1}\NormalTok{, }\DataTypeTok{prob=}\FloatTok{0.92}\NormalTok{)}
\end{Highlighting}
\end{Shaded}

\noindent In Section 2.3 of AGS, it is stated that ``for each patient, whatever the population, occurrence of the event of interest is governed by a proportional hazard model with a constant hazard'', such that:
\begin{equation*}
h = h_0 \times \textnormal{exp}\left((b_A \times (\textnormal{``study A''}) + b_B \times (\textnormal{``study B''})) \times \textnormal{arm} + b_1 \times \textnormal{Refr} + b_2 \times \textnormal{ISS} + b_3 \times \textnormal{PLNEN}\right),
\label{equation1}
\end{equation*}
where $h_0$ denotes the baseline hazard function, assumed constant; and $b_1$, $b_2$, and $b_3$ represent the conditional prognostic effects of Refr (a binary variable indicating ``refractory to some drug class''), ISS and PLNEN (the number of prior treatment lines), respectively. The true \textit{conditional} treatment effect (log hazard ratio) for $A$ vs. $C$ (in both $S=1$ and $S=2$) is $b_A=\log(0.53)$ and the true \textit{conditional} treatment effect for $B$ vs. $C$ in both study populations is $b_B = \log(0.55)$. Note that, due to the non-collapsibility of the (log) hazard ratio, these conditional effects are specific to the adjustment set of covariates used in the outcome-generating process. AGS mention that latent event times for both studies, without accounting for censoring, follow an exponential distribution but do not provide a value for the rate parameter. We set this to $0.5/365$. We use the data-generating process of Bender et al.\cite{bender2005generating}~to simulate exponentially-distributed survival times under proportional hazards.

\begin{Shaded}
\begin{Highlighting}[]
\CommentTok{\# outcome model parameters}
\NormalTok{b\_A <{-}}\StringTok{ }\KeywordTok{log}\NormalTok{(}\FloatTok{0.53}\NormalTok{) }\CommentTok{\# conditional treatment effect (log HR) for A vs. C in study A}
\NormalTok{b\_B <{-}}\StringTok{ }\KeywordTok{log}\NormalTok{(}\FloatTok{0.55}\NormalTok{) }\CommentTok{\# conditional treatment effect (log HR) for B vs. C in study B}
\NormalTok{b\_1}\NormalTok{ <{-}}\StringTok{ }\FloatTok{1.0682} \CommentTok{\# conditional effect of covariate Refr}
\NormalTok{b\_2}\NormalTok{ <{-}}\StringTok{ }\FloatTok{{-}0.6651} \CommentTok{\# conditional effect of covariate ISS}
\NormalTok{b\_3}\NormalTok{ <{-}}\StringTok{ }\FloatTok{0.0825} \CommentTok{\# conditional effect of covariate PLNEN}
\NormalTok{rate <{-}}\StringTok{ }\FloatTok{0.5}\OperatorTok{/}\DecValTok{365} \CommentTok{\# rate of latent time distribution (not specified by AGS)}
\NormalTok{cens\_rate <{-}}\StringTok{ }\FloatTok{0.1}\OperatorTok{/}\DecValTok{365} \CommentTok{\# rate of censoring distribution as set by AGS}

\CommentTok{\# latent event times simulated according to Bender et al. (2005)}
\NormalTok{surv.sim <{-}}\StringTok{ }\ControlFlowTok{function}\NormalTok{(N, LP, rate, cens\_rate) \{}
\NormalTok{  U <{-}}\StringTok{ }\KeywordTok{runif}\NormalTok{(}\DataTypeTok{n=}\NormalTok{N)}
\NormalTok{  Tlat <{-}}\StringTok{ }\OperatorTok{{-}}\KeywordTok{log}\NormalTok{(U)}\OperatorTok{/}\NormalTok{(rate}\OperatorTok{*}\KeywordTok{exp}\NormalTok{(LP)) }\CommentTok{\# latent event times (exponential)}
\NormalTok{  C <{-}}\StringTok{ }\KeywordTok{rexp}\NormalTok{(}\DataTypeTok{n=}\NormalTok{N, }\DataTypeTok{rate=}\NormalTok{cens\_rate) }\CommentTok{\# censoring distribution (exponential)}
\NormalTok{  time <{-}}\StringTok{ }\KeywordTok{pmin}\NormalTok{(Tlat, C) }\CommentTok{\# final follow{-}up times}
\NormalTok{  status <{-}}\StringTok{ }\KeywordTok{as.numeric}\NormalTok{(Tlat}\OperatorTok{<=}\NormalTok{C) }\CommentTok{\# final event indicators}
  \KeywordTok{return}\NormalTok{(}\KeywordTok{cbind}\NormalTok{(time, status))}
\NormalTok{\}}

\CommentTok{\# simulate survival times and event indicators under A and C in S=1}
\NormalTok{LP\_AC <{-}}\StringTok{ }\NormalTok{b\_1}\OperatorTok{*}\NormalTok{plnen\_S1 }\OperatorTok{+}\StringTok{ }\NormalTok{b\_2}\OperatorTok{*}\NormalTok{iss\_S1 }\OperatorTok{+}\StringTok{ }\NormalTok{b\_3}\OperatorTok{*}\NormalTok{refr\_S1 }\OperatorTok{+}\StringTok{ }\NormalTok{b\_A}\OperatorTok{*}\NormalTok{trt }\CommentTok{\# linear predictor }
\NormalTok{survival\_AC <{-}}\StringTok{ }\KeywordTok{surv.sim}\NormalTok{(}\DataTypeTok{N=}\NormalTok{N, }\DataTypeTok{LP=}\NormalTok{LP\_AC, }\DataTypeTok{rate=}\NormalTok{rate, }\DataTypeTok{cens\_rate=}\NormalTok{cens\_rate)}
\NormalTok{time\_AC <{-}}\StringTok{ }\NormalTok{survival\_AC[,}\DecValTok{1}\NormalTok{]}
\NormalTok{status\_AC <{-}}\StringTok{ }\NormalTok{survival\_AC[,}\DecValTok{2}\NormalTok{]}

\CommentTok{\# simulate survival times and event indicators under B and C in S=2}
\NormalTok{LP\_BC <{-}}\StringTok{ }\NormalTok{b\_1}\OperatorTok{*}\NormalTok{plnen\_S2 }\OperatorTok{+}\StringTok{ }\NormalTok{b\_2}\OperatorTok{*}\NormalTok{iss\_S2 }\OperatorTok{+}\StringTok{ }\NormalTok{b\_3}\OperatorTok{*}\NormalTok{refr\_S2 }\OperatorTok{+}\StringTok{ }\NormalTok{b\_B}\OperatorTok{*}\NormalTok{trt }
\NormalTok{survival\_BC <{-}}\StringTok{ }\KeywordTok{surv.sim}\NormalTok{(}\DataTypeTok{N=}\NormalTok{N, }\DataTypeTok{LP=}\NormalTok{LP\_BC, }\DataTypeTok{rate=}\NormalTok{rate, }\DataTypeTok{cens\_rate=}\NormalTok{cens\_rate)}
\NormalTok{time\_BC <{-}}\StringTok{ }\NormalTok{survival\_BC[,}\DecValTok{1}\NormalTok{]}
\NormalTok{status\_BC <{-}}\StringTok{ }\NormalTok{survival\_BC[,}\DecValTok{2}\NormalTok{]}
\end{Highlighting}
\end{Shaded}

\subsection*{True marginal and conditional treatment effects for each study}

We calculate the true value of the marginal or population-average treatment effect $\Delta_{AC}^{(1)}$ for $A$ vs.~$C$ in the $S=1$ population, and the true value of the marginal treatment effect $\Delta_{BC}^{(2)}$ for $B$ vs.~$C$ in the $S=2$ population. The marginal effect is the expected difference in the potential outcomes on the log hazard ratio scale if all members of a given population were under active treatment and if all members of the population were under the common comparator. Each simulated population can be conceptualized as the population of a very large randomized experiment or as two potential cohorts of 50,000 subjects, with one cohort under the active treatment and the other under the common comparator. 

As the populations are sufficiently large to minimize sampling variability, the true marginal effects are computed by fitting simple univariable Cox regressions, regressing the simulated survival times on the indicator variable denoting treatment status. The estimated treatment coefficient of each regression represents the average difference in the potential outcomes on the log hazard ratio scale, and serves as the log of the true marginal hazard ratio for the two interventions under consideration. This is because the survival times have been generated according to the true data-generating mechanism of AGS, where the true conditional effects are explicit, and which uses the correct conditional model by definition. Due to the non-collapsibility of the hazard ratio, this simulation-based approach has been adopted in previous research to determine the true marginal effect.\cite{austin2013performance, austin2016variance, lesko2017bias} 

\begin{Shaded}
\begin{Highlighting}[]
\KeywordTok{library}\NormalTok{(}\StringTok{"survival"}\NormalTok{) }\CommentTok{\# to fit Cox proportional hazards models}
\NormalTok{unadjusted\_model\_AC <{-}}\StringTok{ }\KeywordTok{coxph}\NormalTok{(}\KeywordTok{Surv}\NormalTok{(time\_AC, status\_AC)}\OperatorTok{\midtilde{}}\NormalTok{trt) }
\KeywordTok{exp}\NormalTok{(unadjusted\_model\_AC}\OperatorTok{$}\NormalTok{coefficients[}\DecValTok{1}\NormalTok{]) }\CommentTok{\# marginal hazard ratio for A vs. C (S=1)}
\end{Highlighting}
\end{Shaded}

\begin{verbatim}
##       trt 
## 0.7575748
\end{verbatim}

\begin{Shaded}
\begin{Highlighting}[]
\NormalTok{unadjusted\_model\_BC <{-}}\StringTok{ }\KeywordTok{coxph}\NormalTok{(}\KeywordTok{Surv}\NormalTok{(time\_BC, status\_BC)}\OperatorTok{\midtilde{}}\NormalTok{trt) }
\KeywordTok{exp}\NormalTok{(unadjusted\_model\_BC}\OperatorTok{$}\NormalTok{coefficients[}\DecValTok{1}\NormalTok{]) }\CommentTok{\# marginal hazard ratio for B vs. C (S=2)}
\end{Highlighting}
\end{Shaded}

\begin{verbatim}
##       trt 
## 0.7697989
\end{verbatim}

\noindent We know the true conditional treatment effects for $A$ vs.~$C$ in $S=1$
and for $B$ vs.~$C$ in $S=2$ to be $b_A=\log(0.53)$ and $b_B=\log(0.55)$, respectively. Indeed, these values can be recovered by fitting multivariable regressions of outcome on treatment and the covariates to the simulated data.  

\begin{Shaded}
\begin{Highlighting}[]
\NormalTok{adjusted\_model\_AC <{-}}\StringTok{ }\KeywordTok{coxph}\NormalTok{(}\KeywordTok{Surv}\NormalTok{(time\_AC, status\_AC)}\OperatorTok{\midtilde{}{}}\NormalTok{trt}\OperatorTok{+}\NormalTok{plnen\_S1}\OperatorTok{+}\NormalTok{iss\_S1}\OperatorTok{+}\NormalTok{refr\_S1)}
\KeywordTok{exp}\NormalTok{(adjusted\_model\_AC}\OperatorTok{$}\NormalTok{coefficients[}\DecValTok{1}\NormalTok{]) }\CommentTok{\# conditional hazard ratio for A vs. C (S=1)}
\end{Highlighting}
\end{Shaded}

\begin{verbatim}
##       trt 
## 0.5294677
\end{verbatim}

\begin{Shaded}
\begin{Highlighting}[]
\NormalTok{adjusted\_model\_BC <{-}}\StringTok{ }\KeywordTok{coxph}\NormalTok{(}\KeywordTok{Surv}\NormalTok{(time\_BC, status\_BC)}\OperatorTok{\midtilde{}}\NormalTok{trt}\OperatorTok{+}\NormalTok{plnen\_S2}\OperatorTok{+}\NormalTok{iss\_S2}\OperatorTok{+}\NormalTok{refr\_S2)}
\KeywordTok{exp}\NormalTok{(adjusted\_model\_BC}\OperatorTok{$}\NormalTok{coefficients[}\DecValTok{1}\NormalTok{]) }\CommentTok{\# conditional hazard ratio for B vs. C (S=2)}
\end{Highlighting}
\end{Shaded}

\begin{verbatim}
##       trt 
## 0.5500948
\end{verbatim}

\noindent As expected, conditional and marginal hazard ratios do not coincide, due to the non-collapsibility of the hazard ratio. With non-collapsible measures of effect, conditional treatment effects will vary depending on the covariates used for adjustment in the regression and on the model specification. This may explain why the mean conditional hazard ratios in Table 1 and Table 2 of AGS overestimate $b_A$ and $b_B$, respectively. This is not bias but a by-product of using a different adjustment set where the measure of effect is non-collapsible. 

\subsection*{MAIC analyses}

We now conduct MAIC, estimating the marginal treatment effect for $A$ vs.$C$ in the $S=2$ population. We consider the selection of
the balancing set of covariates.

\begin{Shaded}
\begin{Highlighting}[]
\CommentTok{\# center the S=1 covariates on the S=2 means}
\NormalTok{cent\_age\_S1 <{-}}\StringTok{ }\NormalTok{age\_S1 }\OperatorTok{{-}}\StringTok{ }\KeywordTok{mean}\NormalTok{(age\_S2)}
\NormalTok{cent\_plnen\_S1 <{-}}\StringTok{ }\NormalTok{plnen\_S1 }\OperatorTok{{-}}\StringTok{ }\KeywordTok{mean}\NormalTok{(plnen\_S2)}
\NormalTok{cent\_iss\_S1 <{-}}\StringTok{ }\NormalTok{iss\_S1 }\OperatorTok{{-}}\StringTok{ }\KeywordTok{mean}\NormalTok{(iss\_S2)}
\NormalTok{cent\_refr\_S1 <{-}}\StringTok{ }\NormalTok{refr\_S1 }\OperatorTok{{-}}\StringTok{ }\KeywordTok{mean}\NormalTok{(refr\_S2)}

\CommentTok{\# MAIC function}
\NormalTok{maic <{-}}\StringTok{ }\ControlFlowTok{function}\NormalTok{(X) \{ }\CommentTok{\# X: centered S=1 covariates}
  \CommentTok{\# objective function to be minimized for weight estimation}
\NormalTok{  Q <{-}}\StringTok{ }\ControlFlowTok{function}\NormalTok{(alpha, X) \{}
    \KeywordTok{return}\NormalTok{(}\KeywordTok{sum}\NormalTok{(}\KeywordTok{exp}\NormalTok{(X }\OperatorTok{\%*\%}\StringTok{ }\NormalTok{alpha)))}
\NormalTok{  \}}
\NormalTok{  X <{-}}\StringTok{ }\KeywordTok{as.matrix}\NormalTok{(X)}
\NormalTok{  N <{-}}\StringTok{ }\KeywordTok{nrow}\NormalTok{(X)}
\NormalTok{  K <{-}}\StringTok{ }\KeywordTok{ncol}\NormalTok{(X)}
\NormalTok{  alpha <{-}}\StringTok{ }\KeywordTok{rep}\NormalTok{(}\DecValTok{1}\NormalTok{,K) }\CommentTok{\# arbitrary starting point for the optimizer}
  \CommentTok{\# objective function minimized using BFGS}
\NormalTok{  Q.min <{-}}\StringTok{ }\KeywordTok{optim}\NormalTok{(}\DataTypeTok{fn=}\NormalTok{Q, }\DataTypeTok{X=}\NormalTok{X, }\DataTypeTok{par=}\NormalTok{alpha, }\DataTypeTok{method=}\StringTok{"BFGS"}\NormalTok{)}
\NormalTok{  hat.alpha <{-}}\StringTok{ }\NormalTok{Q.min}\OperatorTok{$}\NormalTok{par }\CommentTok{\# finite solution is the logistic regression parameters}
\NormalTok{  log.hat.w <{-}}\StringTok{ }\KeywordTok{rep}\NormalTok{(}\DecValTok{0}\NormalTok{, N)}
  \ControlFlowTok{for}\NormalTok{ (k }\ControlFlowTok{in} \DecValTok{1}\OperatorTok{:}\NormalTok{K) \{}
\NormalTok{    log.hat.w <{-}}\StringTok{ }\NormalTok{log.hat.w }\OperatorTok{+}\StringTok{ }\NormalTok{hat.alpha[k]}\OperatorTok{*}\NormalTok{X[,k]}
\NormalTok{  \}}
\NormalTok{  hat.w <{-}}\StringTok{ }\KeywordTok{exp}\NormalTok{(log.hat.w) }\CommentTok{\# estimated weights }
\NormalTok{\}}
\end{Highlighting}
\end{Shaded}

\noindent In the first scenario, Scenario 1, we consider balancing the prognostic variables PLNEN, Refr and ISS. According to AGS, this is the correct set of covariates. 

\begin{Shaded}
\begin{Highlighting}[]
\CommentTok{\# centered covariates}
\NormalTok{cent\_X\_S1\_prognostic <{-}}\StringTok{ }\KeywordTok{cbind}\NormalTok{(cent\_plnen\_S1, cent\_iss\_S1, cent\_refr\_S1)}
\NormalTok{weights\_prognostic <{-}}\StringTok{ }\KeywordTok{maic}\NormalTok{(cent\_X\_S1\_prognostic)}
\CommentTok{\# fit weighted Cox proportional hazards model, robust=TRUE for robust sandwich variance}
\NormalTok{maic\_outcome\_model\_1}\NormalTok{ <{-}}\StringTok{ }\KeywordTok{coxph}\NormalTok{(}\KeywordTok{Surv}\NormalTok{(time\_AC, status\_AC)}\OperatorTok{\midtilde{}}\NormalTok{trt, }\DataTypeTok{robust=}\OtherTok{TRUE}\NormalTok{, }
                              \DataTypeTok{weights=}\NormalTok{weights\_prognostic)}
\CommentTok{\# marginal hazard ratio for A vs. C in S=2 population}
\KeywordTok{exp}\NormalTok{(maic\_outcome\_model\_1}\OperatorTok{$}\NormalTok{coefficients)}
\end{Highlighting}
\end{Shaded}

\begin{verbatim}
##       trt 
## 0.7575572
\end{verbatim}

\noindent The marginal or population-average log hazard ratio for $A$ vs.~$C$ in the $S=2$ population ($\Delta_{AC}^{(2)}=\log(0.76)$) is virtually identical to that in the $S=1$ population ($\Delta_{AC}^{(1)}=\log(0.76)$). There is no bias to remove by MAIC as there is no effect modification. Note that we have assumed that MAIC estimates are unbiased under no failures of assumptions, as recently shown by a simulation study in the context of survival outcomes and the Cox proportional hazards model.\cite{remiro2020methods}  

In a second scenario, Scenario 2, we consider balancing a sparse set of covariates. We only balance PLNEN. This selection is questionable because: (1) PLNEN is already balanced; and (2) it has the lowest explanatory power of the prognostic variables --- according to AGS, it is the prognostic covariate most often discarded by LASSO.

\begin{Shaded}
\begin{Highlighting}[]
\NormalTok{cent\_X\_S1\_sparse <{-}}\StringTok{ }\NormalTok{cent\_plnen\_S1}
\NormalTok{weights\_sparse <{-}}\StringTok{ }\KeywordTok{maic}\NormalTok{(cent\_X\_S1\_sparse)}
\NormalTok{maic\_outcome\_model\_2}\NormalTok{ <{-}}\StringTok{ }\KeywordTok{coxph}\NormalTok{(}\KeywordTok{Surv}\NormalTok{(time\_AC, status\_AC)}\OperatorTok{\midtilde{}}\NormalTok{trt, }\DataTypeTok{robust=}\OtherTok{TRUE}\NormalTok{, }
                              \DataTypeTok{weights=}\NormalTok{weights\_sparse)}
\CommentTok{\# marginal hazard ratio for A vs. C in S=2 population}
\KeywordTok{exp}\NormalTok{(maic\_outcome\_model\_2}\OperatorTok{$}\NormalTok{coefficients)}
\end{Highlighting}
\end{Shaded}

\begin{verbatim}
##       trt 
## 0.7575059
\end{verbatim}

\noindent Again, the marginal or population-average hazard ratio for $A$ vs.~$C$ in the $S=2$ population is practically equal to that in the $S=1$ population. There is no bias to remove because there are no treatment effect modifiers in imbalance. Hence, only precision and not bias drives accuracy in this setup. Accounting for a lesser number of covariates will always result in greater accuracy or efficiency due to lower reductions in effective sample size and greater precision, i.e., decreased standard error. This explains why the LASSO approach is more efficient than selecting the maximal set of covariates in the article by AGS.

AGS report relative bias for each simulation scenario in Tables 5 to 9 of the original article. In this particular simulation study, the observed relative bias may arise due to a conflation of marginal and conditional measures of effect. For instance, in our Scenario 1, we can compute the marginal hazard ratio for $A$ vs.~$B$ in the $S=2$ population by dividing the marginal hazard ratio for $A$ vs.~$C$ in $S=2$ by the marginal hazard ratio for $B$ vs.~$C$ in $S=2$.  

\begin{Shaded}
\begin{Highlighting}[]
\KeywordTok{exp}\NormalTok{(maic\_outcome\_model\_1}\OperatorTok{$}\NormalTok{coefficients)}\OperatorTok{/}\KeywordTok{exp}\NormalTok{(unadjusted\_model\_BC}\OperatorTok{$}\NormalTok{coefficients[}\DecValTok{1}\NormalTok{])}
\end{Highlighting}
\end{Shaded}

\begin{verbatim}
##       trt 
## 0.9840976
\end{verbatim}

\noindent This does not coincide with the true estimand defined by AGS, which is a ratio of conditional hazard ratios, $\exp(b_A)/\exp(b_B)=0.53/0.55=0.964$. The relative bias observed by AGS is likely due to this conflation of effects and the non-collapsibility of the hazard ratio.

\subsection*{MAIC analyses with effect modification}

We now consider a scenario, Scenario 3, where Age is a treatment effect modifier for $A$ vs.~$C$ (and also $B$ vs.~$C$) on the (additive) log hazard ratio scale. Interaction terms with coefficient $b_{int}=0.005$ are added to the linear predictor in the outcome-generating model for both studies. 
\begin{Shaded}
\begin{Highlighting}[]
\NormalTok{b\_int <{-}}\StringTok{ }\FloatTok{0.005} \CommentTok{\# interaction coefficient for Age}
\CommentTok{\# simulate outcomes under treatments A and C in S=1}
\NormalTok{LP\_AC =}\StringTok{ }\NormalTok{b\_1}\OperatorTok{*}\NormalTok{plnen\_S1 }\OperatorTok{+}\StringTok{ }\NormalTok{b\_2}\OperatorTok{*}\NormalTok{iss\_S1 }\OperatorTok{+}\StringTok{ }\NormalTok{b\_3}\OperatorTok{*}\NormalTok{refr\_S1 }\OperatorTok{+}\StringTok{ }\NormalTok{b\_A}\OperatorTok{*}\NormalTok{trt }\OperatorTok{+}\StringTok{ }\NormalTok{b\_int}\OperatorTok{*}\NormalTok{age\_S1}\OperatorTok{*}\NormalTok{trt}
\NormalTok{survival\_AC <{-}}\StringTok{ }\KeywordTok{surv.sim}\NormalTok{(}\DataTypeTok{N=}\NormalTok{N, }\DataTypeTok{LP=}\NormalTok{LP\_AC, }\DataTypeTok{rate=}\NormalTok{rate, }\DataTypeTok{cens\_rate=}\NormalTok{cens\_rate)}
\NormalTok{time\_AC <{-}}\StringTok{ }\NormalTok{survival\_AC[,}\DecValTok{1}\NormalTok{]}
\NormalTok{status\_AC <{-}}\StringTok{ }\NormalTok{survival\_AC[,}\DecValTok{2}\NormalTok{]}

\CommentTok{\# simulate outcomes under treatments B and C in S=2}
\NormalTok{LP\_BC <{-}}\StringTok{ }\NormalTok{b\_1}\OperatorTok{*}\NormalTok{plnen\_S2 }\OperatorTok{+}\StringTok{ }\NormalTok{b\_2}\OperatorTok{*}\NormalTok{iss\_S2 }\OperatorTok{+}\StringTok{ }\NormalTok{b\_3}\OperatorTok{*}\NormalTok{refr\_S2 }\OperatorTok{+}\StringTok{ }\NormalTok{b\_B}\OperatorTok{*}\NormalTok{trt }\OperatorTok{+}\StringTok{ }\NormalTok{b\_int}\OperatorTok{*}\NormalTok{age\_S2}\OperatorTok{*}\NormalTok{trt}
\NormalTok{survival\_BC <{-}}\StringTok{ }\KeywordTok{surv.sim}\NormalTok{(}\DataTypeTok{N=}\NormalTok{N, }\DataTypeTok{LP=}\NormalTok{LP\_BC, }\DataTypeTok{rate=}\NormalTok{rate, }\DataTypeTok{cens\_rate=}\NormalTok{cens\_rate)}
\NormalTok{time\_BC <{-}}\StringTok{ }\NormalTok{survival\_BC[,}\DecValTok{1}\NormalTok{]}
\NormalTok{status\_BC <{-}}\StringTok{ }\NormalTok{survival\_BC[,}\DecValTok{2}\NormalTok{]}
\end{Highlighting}
\end{Shaded}

\noindent We consider balancing the maximal set of covariates and use MAIC to estimate the
marginal treatment effect for $A$ vs.~$C$ in the $S=2$ population.

\begin{Shaded}
\begin{Highlighting}[]
\CommentTok{\# centered covariates}
\NormalTok{cent\_X\_S1\_interactions <{-}}\StringTok{ }\KeywordTok{cbind}\NormalTok{(cent\_plnen\_S1, cent\_iss\_S1, cent\_refr\_S1, cent\_age\_S1)}
\NormalTok{weights\_interactions <{-}}\StringTok{ }\KeywordTok{maic}\NormalTok{(cent\_X\_S1\_interactions)}
\CommentTok{\# fit weighted Cox proportional hazards model, robust=TRUE for robust sandwich variance}
\NormalTok{maic\_outcome\_model\_3}\NormalTok{ <{-}}\StringTok{ }\KeywordTok{coxph}\NormalTok{(}\KeywordTok{Surv}\NormalTok{(time\_AC, status\_AC)}\OperatorTok{\midtilde{}}\NormalTok{trt, }\DataTypeTok{robust=}\OtherTok{TRUE}\NormalTok{, }
                              \DataTypeTok{weights=}\NormalTok{weights\_interactions)}
\CommentTok{\# marginal hazard ratio for A vs. C in S=2 population}
\KeywordTok{exp}\NormalTok{(maic\_outcome\_model\_3}\OperatorTok{$}\NormalTok{coefficients[}\DecValTok{1}\NormalTok{])}
\end{Highlighting}
\end{Shaded}

\begin{verbatim}
##       trt 
## 0.8765244
\end{verbatim}

\noindent Because there is a treatment effect modifier, Age, that is in imbalance between the populations, the marginal hazard ratio for $A$ vs.~$C$ in $S=2$ differs from that in $S=1$. A virtually identical marginal hazard ratio for $A$ vs.~$C$ in $S=2$ is estimated if we only balance Age (Scenario 4), such that balancing the other covariates makes no difference in terms of bias reduction.

\begin{Shaded}
\begin{Highlighting}[]
\NormalTok{cent\_X\_S1\_age <{-}}\StringTok{ }\NormalTok{cent\_age\_S1}
\NormalTok{weights\_age <{-}}\StringTok{ }\KeywordTok{maic}\NormalTok{(cent\_X\_S1\_age)}
\NormalTok{maic\_outcome\_model\_4}\NormalTok{ <{-}}\StringTok{ }\KeywordTok{coxph}\NormalTok{(}\KeywordTok{Surv}\NormalTok{(time\_AC, status\_AC)}\OperatorTok{\midtilde{}}\NormalTok{trt, }\DataTypeTok{robust=}\OtherTok{TRUE}\NormalTok{, }
                              \DataTypeTok{weights=}\NormalTok{weights\_age)}
\KeywordTok{exp}\NormalTok{(maic\_outcome\_model\_4}\OperatorTok{$}\NormalTok{coefficients[}\DecValTok{1}\NormalTok{])}
\end{Highlighting}
\end{Shaded}

\begin{verbatim}
##       trt 
## 0.8769922
\end{verbatim}

\section*{Acknowledgments}

We thank the reviewers and the editors of Statistics in Medicine for their constructive feedback, which has been extremely valuable and insightful, and helped in substantially improving our work. Peer Reviewer 1 of a previous article of the authors,\cite{remiro2020methods} helped improve this article further with comments on effect modification and interaction in the Cox model. This article is based on research supported by Antonio Remiro-Az\'ocar's PhD scholarship from the Engineering and Physical Sciences Research Council of the United Kingdom. Anna Heath is supported by the Canada Research Chair in Statistical Trial Design and funded by the Discovery Grant Program of the Natural Sciences and Engineering Research Council of Canada (RGPIN-2021-03366).

\subsection*{Financial disclosure}

Funding agreements ensure the authors' independence in writing and publishing this article. 

\subsection*{Conflict of interest}

The authors declare no potential conflicts of interest.

\subsection*{Data Availability Statement}

The code required to reproduce the analyses is available in the Appendix. 

\bibliography{wileyNJD-AMA}

\end{document}